\documentclass[leqno,12pt]{article}
\usepackage{cite}
\usepackage[a4paper,hscale=0.82,vscale=0.75]{geometry}
\usepackage{graphicx}
\usepackage{amssymb}

\usepackage{amsmath}
\usepackage{amsthm}
\usepackage{amscd}
\usepackage[all,cmtip]{xy} 
\numberwithin{equation}{section}
\usepackage{empheq}
\usepackage[british]{babel}
\usepackage[small,nohug]{diagrams} 
\usepackage{enumitem}
\setlist[enumerate]{label=\arabic*$^\circ$,leftmargin=*,before=\setlength{\rightmargin}{\leftmargin}}

\title{The implications of Galilean invariance\\
for classical point particle lagrangians}
\author{Ziyang Hu\footnote{\texttt{z.hu@damtp.cam.ac.uk}}\\
D.A.M.T.P.\\
 University of Cambridge}

\newcommand{\pd}{\partial}

\DeclareMathOperator{\inp}{\lrcorner}

\newtheoremstyle{shape0}
  {9pt}
  {9pt}
  {}
  {}
  {\bfseries}
  {.}
  {.5em}
  {}

\newtheoremstyle{shape1}
  {9pt}
  {9pt}
  {}
  {\parindent}
  {\it}
  {.}
  { }
  {}

\newtheoremstyle{shape2}
  {9pt}
  {9pt}
  {}
  {}
  {\itshape}
  {.}
  {.5em}
  {}

\theoremstyle{shape1}

\theoremstyle{shape0}

\theoremstyle{shape2}
\theoremstyle{definition}

\begin{document}
\maketitle
\abstract{
We explore the implications of the requirement of Galilean invariance for classical point particle lagrangians, in which the space is not assumed to be flat to begin with. We show that for the free, time-independent lagrangian, this requirement 
is equivalent to the existence of gradient Killing vectors on space, which is in turn equivalent to the condition that the space is a direct product, which is totally flat in the Galilean invariant direction. We then consider more general cases and see that there is no simple generalisation to these cases.}

\tableofcontents
\newpage

\section{Introduction}

Galilean invariance is an experimental fact from low energy non-quantum mechanics. From the space-time persepctive, the action of the symmetry group of mechanics on space in this regime is the linear action of the Galilean group:
\[
\begin{pmatrix}
  1&0&\mathbf{0}\\
  t_{0}&1&\mathbf{0}\\
  \mathbf{x}_{0}&\mathbf{v}&R
\end{pmatrix}
\begin{pmatrix}
  1\\
  t\\
  \mathbf{x}
\end{pmatrix}
\]
where $R$ is an antisymmetric matrix, where $t_{0}$ corresponds to the time translation, $\mathbf{x}_{0}$ corresponds to the space translation, $R$ corresponds to the spatial rotation, and $\mathbf{v}$ corresponds to the Galilean boost symmetry. As shown in \cite{cartan-newtonian,cartan-n-t}, this symmetry group can be gauged, and as a consequence we obtain spaces with a Galilean connection, of which a special kind, namely the kind there is a further constraint on the curvature corresponding to the boost, describes exactly Newtonian gravity.

Another perspective, and perhaps a perspective closer to experimental observations, is that Galilean symmetry is simply a Noether symmetry of the free point particle lagrangian  
\[
L=\tfrac{1}{2}(\dot{\mathbf{x}})^{2}
\]
under
\[
\delta x^{i}=v^{i}t,\qquad\delta \dot x^{i}= \delta v^{i},
\]
where $v^{i}$ is infinitesimal. As any Noether symmetry of the lagrangian, it has its conserved current behaving as the ``center of energy'' of the system, which is hardly ever described or used in the standard treatment.

An interesting question about Galilean symmetry arises in relation to Newtonian cosmology \cite{newton}, used, for example, for the study of Friedmann equations, is the following: in Newtonian cosmology we obviously do not want require that the space is flat, as otherwise we will just be returning to the usual case, with perhaps velocity-dependent potential. But what is the status of Galilean symmetry in this setting? Can we treat it as, instead of a ``broken'' symmetry, an unbroken symmetry still? In this paper we shall explore this question, and show that in the time-independent free particle case, the requirement of Galilean symmetry implies decoupled flat directions: in particular, if we insist on the full set of Galilean symmetries we are back in flat space. Then we briefly consider other cases where we add various terms to the lagrangian.

\section{Galilean symmetry for time-independent free system}

Our starting point is the point particle lagragian
\[
L=\frac{1}{2}g_{ij}(x)\dot x^{i}\dot x^{j}.
\]
where the metric $g_{ij}$ is no longer assumed to be flat.
Under the transformation
\[
\delta x^{i}=f^{i}(x,t),\qquad\delta\dot x^{i}=\pd_{t} f^{i}+f^{i}{}_{,k}\dot x^{k},
\]
we have
\[
\delta{L}=\tfrac{1}{2}(g_{ij,k}f^{k}+g_{kj}f^{k}{}_{,i}+g_{ik}f^{k}{}_{,j})\dot x^{i}\dot x^{j}+g_{ij}\pd_{t}f^{i}\dot x^{j}=\frac{1}{2}(\mathcal{L}_{f}g_{ij})\dot x^{i}\dot x^{j}+g_{ij}\pd_{t}f^{i}\dot x^{j}.
\]
This is a candidate for a ``Galilean symmetry'' of the system.
This change is a symmetry of the system if it is a total time derivative of a function $F(x,t)$
\[
\delta L=\frac{dF}{dt}=\pd_{t}F+F_{,k}\dot x^{k}.
\]
Now, as a basic assumption of the lagrangian treatment, the velocity can be taken to be an independent variable. Hence the coefficients of the powers of $\dot x$ must be equal. Therefore
\[
\left\{
  \begin{aligned}
    \pd_{t}F&=0,\\
    F_{,i}&=g_{ij}\pd_{t}f^{j},\\
    0&=\mathcal{L}_{f}g_{ij}
  \end{aligned}
\right.
\]
The first equation means that $F$ is a function of space only: $F=F(x)$, and the third means that at any instant of time, the vector field $f$ is a Killing vector field. So the most general form of $f$ is
\[
f^{i}=\varphi_{a}(t)v^{i}_{a}(x)
\]
where the  $v^{i}_{a}$ form a basis of the Killing vector fields, $a=1,2,\dots k$ where $k$ is the number of Killing vector fields on the space (might as well be zero), and they depend on space only. Then the second equation gives us
\[
\varphi_{a}'(t)v_{ai}(x)=F_{,i}(x)
\]
and we immediately have:
\[
\varphi_{a}(t)=a_{a}t+b_{a},
\]
affine functions of $t$. \footnote{The case $a_a=0$ is the trivial case: the symmetry does not involve time and the Lagrangian only has the spatial symmetries, not the boost symmetries.} But then $a_{a}v_{ai}$ itself is a Killing vector, since now $a_{a}$ are constants. Let us write $v_{i}=a_{a}v_{ai}$, then we also have
\[
v_{[i,j]}=0.
\] 
Hence the system admits a certain number of Galilean boost invariance if and only if the system has the same number of \emph{gradient Killing vectors}. To find such a symmetry is to find a vector field $v^{i}$ such that, first, it is Killing, and second, its covariant form is given by a total derivative. This problem can now be studied completely within the geometry of the space with metric $g_{ij}$ and without consideration of the effects of time.

\section{Gradient Killing vectors}
\label{sec:from-reduction-point}

It is easy to check that in flat space, only the translation Killing vector fields are gradient Killing vector fields, and the only homogeneous space with positive definite metric that admits gradient Killing vector fields at all is Euclidean space. We would like to investigate the problem in the case of a space with a general positive definite Riemannian metric.

Whenever we have Killing vectors, it is best to study the problem by working in a principal bundle in which the number of free group parameters is reduced as a consequence of the symmetry: we will use the framework developed in \cite{hu1} for our present study.
Let us now set up orthonormal moving frames on the Riemannian space with metric $g_{ij}$. If the space is $n$ dimensional, let us align $\omega^{\star}$ along $v_{i}$:
\[
e^{\lambda(x)}\omega^{\star}= v_{i}dx^{i}
\]
where $\exp{\lambda}$ is a scalar function depending on position and is fixed up to a global constant by the orthonormal condition of the moving frame and the norm of the Killing vector field at different points. The quantity $\exp{\lambda}$ is required to be non-negative everywhere, and positive almost everywhere, and hence we write it in this exponential form. The rest of the moving frame is $\omega^{i}$ where now $i=1,2,\dots,n-1$ and
\[
ds^{2}=\omega^{\star}\otimes\omega^{\star}+\sum_{i=1}^{n-1}\omega^{i}\otimes\omega^{i},
\]
which leads us to the formalism of a rigid flow. We have the first Maurer-Cartan relations
\begin{align*}
  d\omega^{\star}&=-K_{i}\omega^{\star}\wedge\omega^{i}-M_{ij}\omega^{i}\wedge\omega^{j},\\
  d\omega^{i}&=-(\omega_{ij}-M_{ij}\omega^{\star})\wedge\omega^{j}
\end{align*}
and as usual, $M_{ij}=-M_{ji}$. The detailed derivation of these equations, valid for much more general systems, in given in \cite{hu1}.

The relation between the isometry given by $v^{i}$ in the total space, which we will now denote by $M$, and the reduced space, which we denote by $B$, is that
\[
\mathbf{I}^{\star}(\lambda)=0,\qquad K_{i}\omega^{i}=d_{B}\lambda,
\]
where $d_{B}$ denotes exterior derivative on the reduce space $B$, and in particular, $K_{i,\star}=0$.
The condition that the Killing vector is locally a gradient vector is, by Poincar\'e's lemma,
\begin{align*}
0&=d(e^{\lambda}\omega^{\star})\\
&=e^{\lambda}(2K_{i}\omega^{i}\wedge\omega^{\star}-M_{ij}\omega^{i}\wedge\omega^{j})
\end{align*}
and hence we are required to have
\[
K_{i}=0,\qquad M_{ij}=0.
\]
in other words, there is now a complete decoupling between the Killing vector field direction and the ``horizontal'' direction, and every space that admits a gradient Killing vector is uniquely determined by giving its reduced space after the reduction, and the total space is simply the product space of the reduced Riemannian space and $\mathbb{E}^{1}$, the Euclidean line, with the product metric on it.

From the discussion above, we now know that if we have a boost symmetry, then we have the frame $\omega^{\star},\omega^{i}$ on the space, and
\begin{align*}
  d\omega^{*}&=0,\\
  d\omega^{i}&=-\omega^{i}{}_{j}\wedge\omega^{j}.
\end{align*}

Now we prove by induction that \emph{the existence of additional gradient Killing vectors implies the flat directions can be extended to totally flat, uncoupled fibrations of higher dimensional}. Assume we already have $\omega^{a}$, $a=1,\dots,k$ the flat directions. The Maurer-Cartan equations are
\begin{align*}
  d\omega^{a}&=0,\\
  d\omega^{i}&=-\omega^{i}{}_{j}\wedge\omega^{j}.
\end{align*}
where now $i=1,2,\dots,n-k$.
Now suppose we have another gradient Killing vector field, which we write as
\[
\mathbf{u}=a^{a}\mathbf{I}_{a}+b^{i}\mathbf{I}_{i}
\]
and with the indeterminacy of an overall constant. We can actually fix this constant: since this Killing vector field must also be a pure translation as  the first one, the vector field has constant norm, and we can set
\[
\sum_{a}(a^{a})^{2}+\sum_{i}(b^{i})^{2}=1.
\]

We now have two conditions that we must impose. First, $\mathbf{u}$ is a Killing vector field, and second, the form
\[
\mu=a^{a}\omega^{a}+b^{i}\omega^{i}
\]
satisfies $d\mu=0$. For the first condition, we need (working with a section)
\begin{align*}
\mathcal{L}_{\mathbf{u}}\omega^{\star}&=da^{a}=a^{a}{}_{,b}\omega^{b}+a^{a}{}_{,i}\omega^{i}\\
\mathcal{L}_{\mathbf{u}}\omega^{i}&=-\mathbf{u}\inp(\omega^{i}{}_{j}\wedge\omega^{j})+db^{i}=b^{i}{}_{,a}\omega^{a}+c_{i}\omega^{i}
\end{align*}
by the magic formula, where $c_{i}$ are functions we do not really care about. Then
\[
\mathcal{L}_{\mathbf{u}}ds^{2}=(a^{a}{}_{,i}+b^{i}{}_{,a})(\omega^{i}\otimes\omega^{a}+\omega^{a}\otimes\omega^{i})+\cdots
\]
where the dots are terms  not containing the cross terms in $\omega^{a}$ and $\omega^{i}$. Hence
\[
a^{a}{}_{,i}+b^{i}{}_{,a}=0.
\]
For the second condition,
\begin{align*}
  0&=d(a^{a}\omega^{a}+b^{i}\omega^{i})\\
  &=(a^{a}{}_{,i}-b^{i}_{,a})\omega^{i}\wedge\omega^{\star}+\cdots
\end{align*}
where again we have omitted terms not containing the cross terms. Hence, we need
\[
a^{a}{}_{,i}-b^{i}{}_{,a}=0,
\]
which together with the earlier condition implies $b^{i}{}_{,a}=0$, $b^{i}$ are functions actually defined on the reduced space.

We also have $a_{a,i}=0$. If we consider now the terms in $\omega^{a}\otimes_{S}\omega^{b}$ and $\omega^{a}\wedge\omega^{b}$ in the above calculations, the conditions we get are
\[
a_{a,b}+a_{b,a}=0,\quad a_{a,b}-a_{b,a}=0\quad\Rightarrow\quad a_{a,b}=0
\]
 which together with the previous condition implies that that $a_{a}$ are constants. Then, using the residual $SO(n-1)$ symmetry for the frames with respect to $\omega^{i}$ and $\mathbf{I}_{i}$, we can write
\[
\mathbf{u}=a_{a}\mathbf{I}_{a}+b\mathbf{I}_{n-k}\\
\] 
where now  $a_{a}$ and $b$ are all constants. Since sums of Killing vector fields with constant coefficients are Killing vector fields themselves, this shows that $\mathbf{I}_{n-k}$ is a Killing vector field, and in this case it is a gradient vector field as well. The Cartan connection matrix  in this frame now becomes
\[
\begin{pmatrix}
  0&0&0&0\\
  \omega^{a}&0&0&0\\
  \omega^{n-k}&0&0&0\\
  \omega^{i}&0&0&\omega^{i}{}_{j}
\end{pmatrix}
\]
where now $i,j=1,2,\dots,n-k-1$. This shows that any $n$ dimensional space with $k$ gradient Killing vector fields can be constructed by a $n-k$ dimensional Riemannian space multiplied with $\mathbb{E}^{k}$, with the direct product metric. In particular, \emph{the only $n$ dimensional space with $n$ gradient Killing vectors is Euclidean space itself.}

\section{More general systems}
\label{sec:gener-time-depend}

There are three ways that we can generalise the Lagrangian that we started with: first, allow a potential term, second, allow space geometry to be time dependent, and third, allow magnetic force terms. The most general Lagrangian can now be written
\[
L=\frac{1}{2}g_{ij}(x,t)\dot x^{i}\dot x^{j}+A_{i}(x,t)\dot x^{i}-V(x,t)
\]
where now $i=1,2,\dots , nk$ where $n$ is the dimension of the manifold the dynamics takes place, and $k$ is the number of particles. As usual,
\[
\delta x^{i}=f^{i}(x,t),\qquad \delta\dot x^{i}=\pd_{t}f^{i}+f^{i}{}_{,k}\dot x^{k}.
\]
Now we have
\[
\delta{L}=\frac{1}{2}(\mathcal{L}_{f}g_{ij})\dot x^{i}\dot x^{j}+g_{ij}\pd_{t}f^{i}\dot x^{j}+(\mathcal{L}_{f}A_{i})\dot x^{i}+A_{i}\pd_{t}f^{i}-V_{,i}f^{i},
\]
as in mechanics, the variation does not involve the term $\delta t$.

The change is a symmetry of the system if it is a total time derivative of a function $F(x,t)$
\[
\delta L=\frac{dF}{dt}=\pd_{t}F+F_{,k}\dot x^{k}.
\]
Now, the velocity is an independent variable. Hence the coefficients of the powers of $\dot x$ must be equal. Therefore
\[
\left\{
  \begin{aligned}
    \pd_{t}F&=A_{i}\pd_{t}f^{i}-V_{,i}f^{i},\\
    F_{,i}&=g_{ij}\pd_{t}f^{j}+\mathcal{L}_{f}A_{i},\\
    0&=\mathcal{L}_{f}g_{ij}
  \end{aligned}
\right.
\]
Again, assume that $v^{i}_{a}$ is a basis of Killing vectors on the space, which now in principle can be time-dependent. For example, locally we can identify a small neighbourhood of Euclidean space with a small neighbourhood of the sphere or the hyperbolic space. Let the deformation of the space be only due to the change of the curvature of the space. Then we see the family of Killing vectors also change smoothly. Actually, in this case, we see that for the canonical coordinates on the sphere, the Euclidean space and the hyperbolic space, the only expression for the Killing vector fields remain the same in a small neighbourhood.

It is in principle also possible that as the space deforms as time goes by there may come up with new Killing vectors, and some of the old Killing vectors may disappear. It is obvious that these ``non-persistent'' Killing vectors will not play a role in what follows and our basis is a basis only for the persistent ones. Then we can write
\[
f^{i}=\varphi_{a}(t)v^{i}_{a}(x,t).
\]

Now let us turn on the three assumptions one by one, to investigate their effects on the problem.

If we only have $V\neq 0$, then our equations become
\[
\left\{
  \begin{aligned}
    f^{i}&=\varphi_{a}(t)v_{a}^{i}(x)\\
    \pd_{t}F&=-V_{,i}\varphi_{a}v_{a}^{i},\\
    F_{,i}&=g_{ij}\varphi'_{a}v^{i}_{a},\\
    0&=\mathcal{L}_{f}g_{ij}.
  \end{aligned}
\right.
\]
Again, at any instant of time, the vector field $\varphi_{a}'v_{a}^{i}$ is a gradient Killing vector field. If the potential is time-dependent, the $\varphi_{a}$ will no longer be an affine function of $t$, but with more complicated time-dependence. If on the other hand $V$ is time-independent as well, $V=V(x)$, then once again $\varphi_{a}$ must be affine functions of $t$. In this case we write $\varphi_{a}=a_{a}t+b_{a}$, and
\[
\left\{
  \begin{aligned}
    \pd_{t}F&=-V_{,i}(a_{a}t+b_{a})v_{a}^{i},\\
    F_{,i}&=g_{ij}a_{a}v^{i}_{a},
  \end{aligned}
\right.
\]
and consistency requires
\[
-\pd_{i}[(a_{a}t+b_{a})(V_{,j}v^{j}_{a})]=0.
\]
If $a_{a}=0$, then we are in the trivial case and no boost symmetry arises. In the non-trivial case, we already have $v^{j}_{a}$ gradient Killing vectors, and by our above arguments in the canonical Cartesian coordinates $\pd_{i}v^{j}_{a}=0$. Therefore we must require, in a canonical Cartesian coordinates,
\[
V_{,ij}=0,
\]
i.e., the potential can be at most linear in the positions. This is the case for the linear gravitational potential on the surface of the earth, but not for the inverse gravitational potential for large bodies. However \emph{this does not really exclude the possibility of boost invariance in these cases}, since under boosts, potentials that are originally time-independent will become time-dependent.

Now consider the case that the metric $g_{ij}$ is time-dependent, without any potential terms. We have

\[
F_{,k}=g_{ij}\pd_{t}(\varphi_{a}v^{i}_{a})=g_{ij}\varphi'_{a}v^{i}_{a}+g_{ij}\varphi_{a}\pd_{t}v^{i}_{a}.
\]
\emph{If the local expression of the Killing vector fields do not change} (it is not true that this can always be upheld by a change of coordinates: in particular, we need to have $\mathcal{L}_{v_{a}}(g_{ij,k}f^{k})=0$, though such coordinates obviously exists for Robertson-Walker style scaling and the above ``mutation'' of homogeneous spaces mentioned above), that is to say, $v^{i}_{a}$ does not depend on $t$, then we return to our previous case
\[
F_{,k}=g_{ij}\varphi'_{a}v^{i}_{a},
\]
and
\[
\left\{
  \begin{aligned}
    f^{i}&=\varphi_{a}(t)v_{a}^{i}(x)\\
    \pd_{t}F&=0,\\
    F_{,i}&=g_{ij}\varphi'_{a}v^{i}_{a},\\
    0&=\mathcal{L}_{f}g_{ij}.
  \end{aligned}
\right.
\]
then for surfaces where $t=\text{constant}$, $\varphi'_{a}v^{i}_{a}$ must still be a gradient Killing vector field. Hence in this case we are back to the business of finding gradient Killing vector fields on a space. The $t$ dependence in $\varphi_{a}(t)$ is in this case no longer affine: it must be chosen to cancel the $t$ dependence in $g_{ij}$ so as to ensure that the partial derivatives on $F$ commutes.

Note that, if we are only interested in finding some Galilean boosts, we can set some of the $\varphi_{a}$ to vanish by hand, which means the vector field $\varphi_{a}'v_{a}^{i}$ depends only on some of the vector fields. Then, as long as these vector fields retain form-invariant under change in time, the above analysis is still valid. For example, the following metric
\[
ds^{2}=\sum_{i}a_{i}(t)dx^{i}dx^{i}
\]
i.e., the different directions are scaled differently, reproduces all the translation Killing vectors if we change variables as time goes by.

Finally, if now we allow $A_{i}$ to be non-zero, then as
\[
F_{,i}=g_{ij}\pd_{t}f^{j}+\mathcal{L}_{f}A_{i},
\]
we are required to have gradient Killing vector fields only if $\mathcal{L}_{f}A_{i}=0$, i.e., $A_{i}$ is constant in the boost directions. If the system is boost-invariant in all directions, then $A_{i}=0$.

\section{Conclusion}

In this paper we have studied the implications of Galilean invariance for point particle lagrangians in which time is treated as absolute, i.e., point particle lagrangians of Newtonian cosmology, and we have shown that, for the time-independent free lagrangian, Galilean invariance requires completely decoupled flat spatial directions. On the other hand, there are no simple generalisations of this result when we add additional terms to the lagrangian.

\bibliographystyle{hplain}

\addcontentsline{toc}{section}{References}
\end{document}